\documentclass[aip,reprint]{revtex4-1}

\usepackage{amsmath}
\usepackage{amssymb}
\usepackage{graphicx}
\usepackage{color}
\usepackage{float}
\usepackage{sistyle}
\usepackage{subfigure}  % use for side-by-side figures
\usepackage{verbatim}   % useful for program listings
\usepackage{graphicx}% Include figure files
\usepackage{dcolumn}% Align table columns on decimal point
\usepackage{bm}% bold math
\usepackage[citecolor=blue,filecolor=blue,colorlinks=true,urlcolor=blue]{hyperref}
\usepackage{ulem}

\begin{document}

\title{Impact of the surface phase transition on magnon and phonon excitations in BiFeO$_3$ nanoparticles}

\author{Ian Aupiais}
\author{Pierre Hemme}
\affiliation{Laboratoire Mat\'eriaux et Ph\'enom$\grave{e}$nes Quantiques (UMR 7162 CNRS), Universit\'e de Paris, Bat. Condorcet, 75205 Paris Cedex 13, France}

\author{Marc Allen}
\affiliation{Department of Physics and Astronomy, University of Victoria, Victoria, British Columbia, Canada V8W 2Y2}
\affiliation{Centre for Advanced Materials and Related Technology, University of Victoria, Victoria, British Columbia, Canada V8W 2Y2}

\author{Alexis Scida}
\affiliation{Department of Chemistry, Oregon State University, 153 Gilbert Hall, Corvallis, Oregon 97331, United States}
	
\author{Xiao Lu}
\affiliation{Department of Chemistry, State University of New York at Stony Brook, Stony Brook, New York 11794, United States}	

\author{Christian Ricolleau}
\author{Yann Gallais}
\author{Alain Sacuto}
\affiliation{Laboratoire Mat\'eriaux et Ph\'enom$\grave{e}$nes Quantiques (UMR 7162 CNRS), Universit\'e de Paris, Bat. Condorcet, 75205 Paris Cedex 13, France}
	
\author{Stanislaus Wong}
\affiliation{Department of Chemistry, State University of New York at Stony Brook, Stony Brook, New York 11794, United States}	

\author{Rogério de Sousa}
\affiliation{Department of Physics and Astronomy, University of Victoria, Victoria, British Columbia, Canada V8W 2Y2}
\affiliation{Centre for Advanced Materials and Related Technology, University of Victoria, Victoria, British Columbia, Canada V8W 2Y2}
	
\author{Maximilien Cazayous}\thanks{corresponding author : maximilien.cazayous@u-paris.fr}
\affiliation{Laboratoire Mat\'eriaux et Ph\'enom$\grave{e}$nes Quantiques (UMR 7162 CNRS), Universit\'e de Paris, Bat. Condorcet, 75205 Paris Cedex 13, France}
	
\begin{abstract}
We have performed Raman scattering measurements on BiFeO$_3$ nanoparticles and studied both magnetic and lattice modes. We reveal strong anomalies between 140 K and 200 K in the frequency of magnon and E(LO$_1$), E(TO$_1$) and A$_1$(LO$_1$) phonon modes. These anomalies are related to a surface expansion and are enhanced for nanoparticle sizes approaching the spin cycloidal length. These observations point out the strong interplay between the surface, the lattice, and the magnetism for sizes of BiFeO$_3$ nanoparticles close to the cycloid periodicity.
\end{abstract}

\maketitle

Bismuth ferrite (BiFeO$_3$) represents a prototypical multiferroic with outstanding properties. This multifunctional material has been associated with a number of diverse technological applications such as photovoltaics \cite{Choi2009,Allibe2010}, electrocatalysis \cite{Mukherjee}, nanoelectronics \cite{Seidel2009,Crassous2011}, energy harvesting \cite{Zeches2009}, and even spintronics \cite{Dho2006,Allibe2012}. Also, it has been demonstrated that controlling the dimensionality of this material allows us to modify its properties. For example, controlling the size of BiFeO$_3$ nanoparticles permits a significant tuning of the magnetic, electrical, and optical properties \cite{Mocherla2013,Carranza2019}. Moreover, it has been proposed that the confinement in BiFeO$_3$ nanoparticles should induce a size-dependent bistability of magnetization and polarization useful for memory bits \cite{DeSousa2019}.

\begin{figure}[h]
	\begin{center}
		\includegraphics[width=8.7cm]{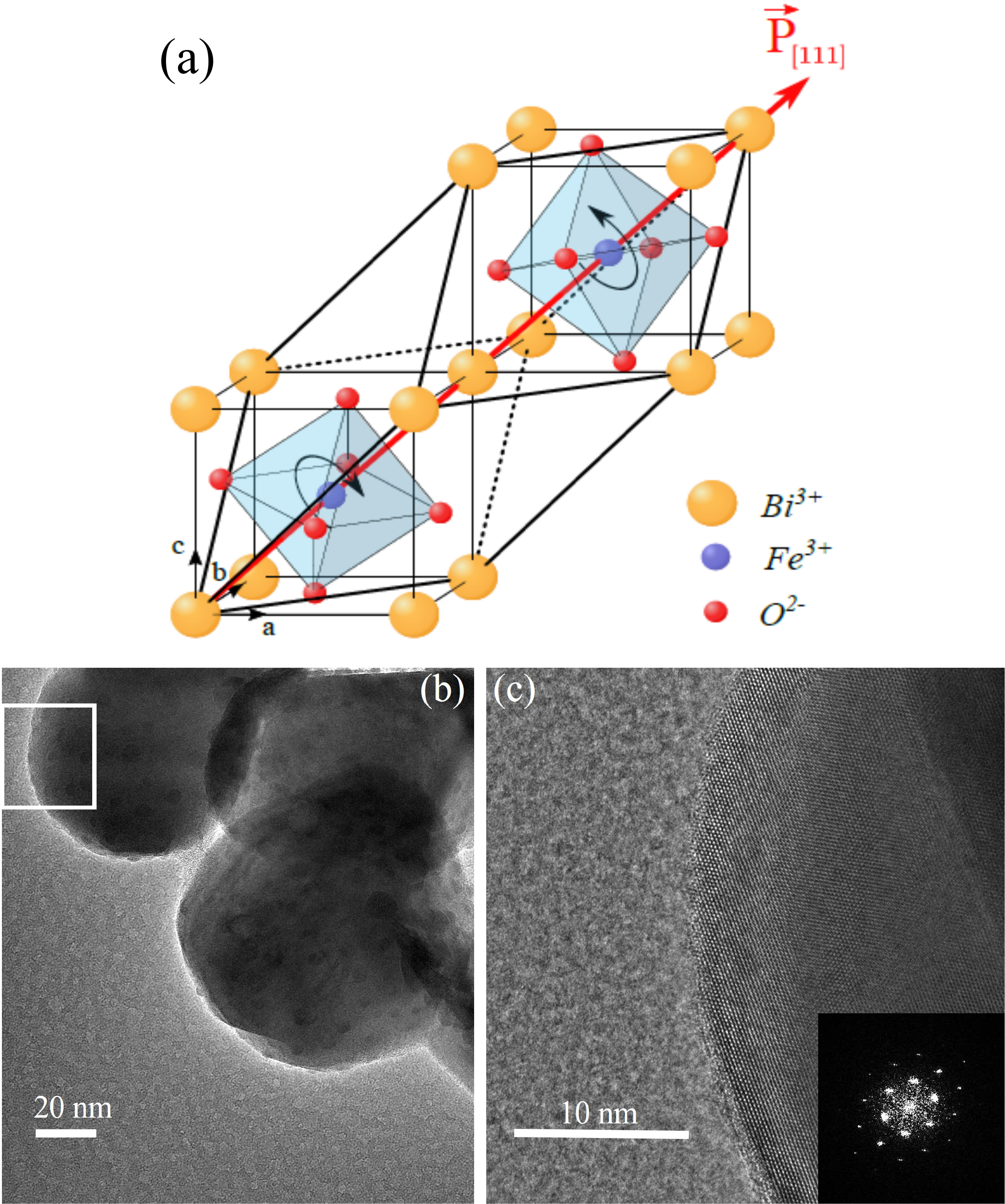}
		\caption{(a) Schematic of the crystal structure of BFO and the ferroelectric polarization (arrow) (b) TEM image of nanoparticles of about 83 nm. (c) Magnification of the marked region in (a) with the associated Fourier transform.}
		\label{fig:0}
	\end{center}
\end{figure}

In bulk phase, BiFeO$_3$ is ferroelectric below the Curie temperature $T_C \sim$ 1100 K with a large spontaneous electrical polarization of $\sim$ 100 $\mu$C$\cdot$cm${^{-2}}$ (see Fig. \ref{fig:0}(a)) \cite{Lebeugle2007, Teague1970}. The %lone-pair
electrons present at the Bi sites are responsible for the cationic displacements %that's breaking the inversion symetry allowing 
at the origin of the ferroelectricity \cite{Neaton2005,Ravindran2006}. In addition, an antiferromagnetic order appears below the Néel temperature, $T_N \sim$ 640 K. This magnetic order corresponds to a G-type antiferromagnetic behavior, modulated by a long-wavelength cycloid of 62 nm \cite{Sosnowska1982}. This modulation is attributed to a homogeneous Dzialoshinskii-Moriya interaction activated by the ferroelectric displacement of the Fe sites \cite{Matsuda2012,Fishman2013,Lee2016}. The cycloid is also slighty tilted out of its propagation plane and characterised by a local ferromagnetic moment cancelled out over a period \cite{Ramazanoglu2011a}. This additional modulation is due to an inhomogeneous Dzialoshinskii-Moriya interaction activated by the rotation of oxygen octahedra \cite{Fishman2013,Lee2016}.
Reducing the size of ferroelectric nanoparticles induces structural distortions which favor the high-symmetry paraelectric phase. In the case of BiFeO$_3$, it has been determined that the paraelectric phase appears below 9-10  nm \cite{Selbach2007,Chen2010}. For antiferromagnetic nanoparticles, the uncompensated antiferromagnetic sublattice at the surface leads to a ferromagnetic moment which increases as the size is reduced \cite{Park2007,AnnapuReddy2012}. Unexpectedly, F. Huang \textit{et al.} have observed an anomalous behavior of the magnetization for a size close to the cycloid periodicity \cite{Huang2013}. In addition, a surface expansion takes place between 140 K and 200 K within the topmost 10 nm of the bulk \cite{Jarrier2012}.

The role of the surface expansion on the dynamical properties of BiFeO$_3$ nanoparticles, for which the surface/volume ratio increases as the size is reduced, remains to be studied. In this work, we have tracked the effect of the surface expansion by the corresponding size reduction on phonons and magnons. We observe the disappearance of the fingerprint of the cycloidal spin excitations between the sizes of 83 nm and 61 nm. At two different temperatures associated with surface expansion, anomalies in the frequency of magnetic and lattice excitations are observed. Such anomalies are enhanced for a nanoparticle size %of 83 nm. In addition, we put in evidence an anomalous behavior associated to the surface expansion which is enhanced at a size 
close to the cycloid periodicity.
%Thus there exists a cross-correlation between the lattice, magnetism and the surface at a size close to the cycloid periodicity.

We performed Raman spectroscopy measurements on BiFeO$_3$ nanoparticles. Five sizes have been synthesized by the sol-gel method which has been adapted and optimized from the protocol reported in Ref. \onlinecite{Lee2016} : 250 nm $\pm$ 100 nm, 158 nm $\pm$ 17 nm, 83 nm $\pm$ 20 nm, 61 nm $\pm$ 9 nm, and 31 nm $\pm$ 7 nm. Figure \ref{fig:0}(b) shows a transmission electron microscopy (TEM) image of nanoparticles with a mean size of 83 nm. The nanoparticles are regularly shaped and approximately spherical. Figure \ref{fig:0}(c) presents a magnification of the region defined in Fig. \ref{fig:0}(b). The Fourier transform  shows the presence of sharp diffraction spots which are indicative of the formation of single crystalline BiFeO$_3$. Platelets of nanoparticles with a thickness of 3 nm have been prepared under very low pressure for Raman measurements. Raman spectra were recorded in a backscattering geometry with a triple spectrometer Jobin Yvon T64000 using the 647.1 nm excitation line from a Ar$^+$-Kr$^+$ mixed gas laser. The penetration depth of the laser light is very large because of the low extinction coefficient $k = 1.2\:10^{-4}$ of BFO \cite{Weber2016}. Thus, BFO is nearly transparent for the 647.1 nm light. The excited volume is given by the focus ($\sim$70 $\mu$m).
 The high rejection rate of the spectrometer allows us to detect excitations as low as 10 cm$^{-1}$. Temperature is controlled with an open cycle He cryostat. Since nanoparticles are randomly oriented, there are no Raman selection rules.  %The Raman signal strongly decreases between a size of 83 nm and 61  nm because of the change of the electronic gap \cite{Mocherla2013}. 
We observe all the phonon modes that characterize the rhombohedral  phase (R3c) for each nanoparticle size. 
% (for more information, see supplementary).

\begin{figure}[h]
	\begin{center}
		\includegraphics[width=8.7cm]{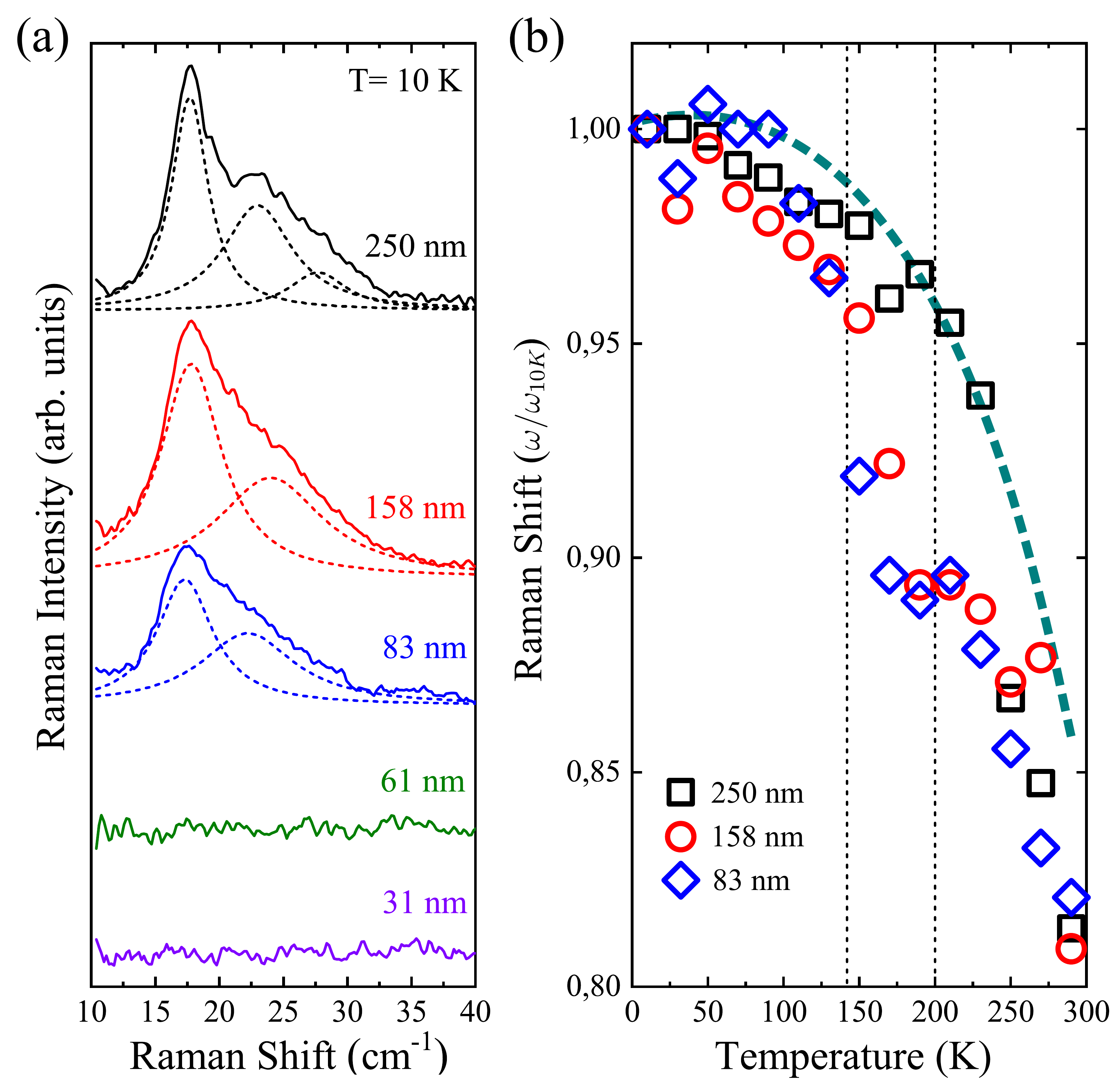}
		\caption{(a) Magnon part of the Raman spectra for the five sizes of nanoparticles. Peaks have been fitted by Lorentzian functions (dashed peaks) (b) Temperature dependence of the magnon frequency with the higher intensity normalised by the value at 10 K for a nanoparticle sizes of 250 nm, 158 nm and 83 nm. Dashed line corresponds to the behavior in the bulk. Vertical dashed lines represent the range where the surface expansion is expected.}
		\label{fig:1}
	\end{center}
\end{figure}

Figure \ref{fig:1}(a) shows the low-energy part of the Raman spectra for the five sizes of nanoparticles. For 250 nm, three peaks are observed. In the bulk phase, the magnetic cycloid manifests itself by two series of narrow peaks in the Raman spectra. They are the fingerprint of the magnon modes \cite{Cazayous2008}. They correspond to the spin excitation oscillations in and out of the cycloidal plane measured at zero wave vector. The peaks observed in Fig. \ref{fig:1}(a) correspond to the superposition of the narrow peaks of the bulk (no Raman selection rules in randomly oriented nanoparticles) and are associated with the cycloidal spin states. For 158 nm and 83 nm, we distinctly only observe two peaks. No peak is detected for sizes of 61 nm and 31 nm. The optical gap of BiFeO$_3$ shifts with the nanoparticle size which could explain the disappearance of the peaks \cite{Mocherla2013}. However, the signature of the cycloid has not been detected using any other laser wavelengths. In Fig. \ref{fig:1}(b), we have reported the frequency of the magnon at 17 cm$^{-1}$ as a function of temperature. For 250 nm, the frequency has a similar dependence as a magnetic excitation in the bulk phase \cite{Cazayous2008}. For 158 nm, the frequency presents a jump around 140 K and a plateau around 200 K. A slightly more pronounced trend is measured for a size of 83 nm. %In the same range of temperature, a surface phase transition with an associated sharp change in lattice parameter and charge density at the surface has been observed.\cite{Mocherla2013}
In the same range of temperature, a surface phase transition with an associated sharp change in lattice parameter at the surface has been observed \cite{Jarrier2012}. This “skin” effect is estimated to be within the topmost 10nm of BFO \cite{Jarrier2012}. For a nanoparticle size of 80 nm, this effect represents 33\% of the volume of the nanoparticle. The observed anomalies in the magnon frequencies of nanoparticles can be attributed to the surface expansion. The effect of this expansion on the spin excitations is enhanced around 83 nm. Hence, our results show that the surface has a significant impact on the magnetism for a size approaching the cycloid periodicity of 62 nm. It is known that BiFeO$_3$ shows a strong magneto-striction and spin-phonon coupling \cite{Matsuda2012, Fishman2013, Arnold2009,Lee2015}. Both phenomena might be at the origin of the observed anomalies. This raises the question of the impact of the surface expansion in nanoparticles on the lattice excitations when the size approaches the cycloid periodicity.

\begin{figure}[h]
	\begin{center}
		\includegraphics[width=8.7cm]{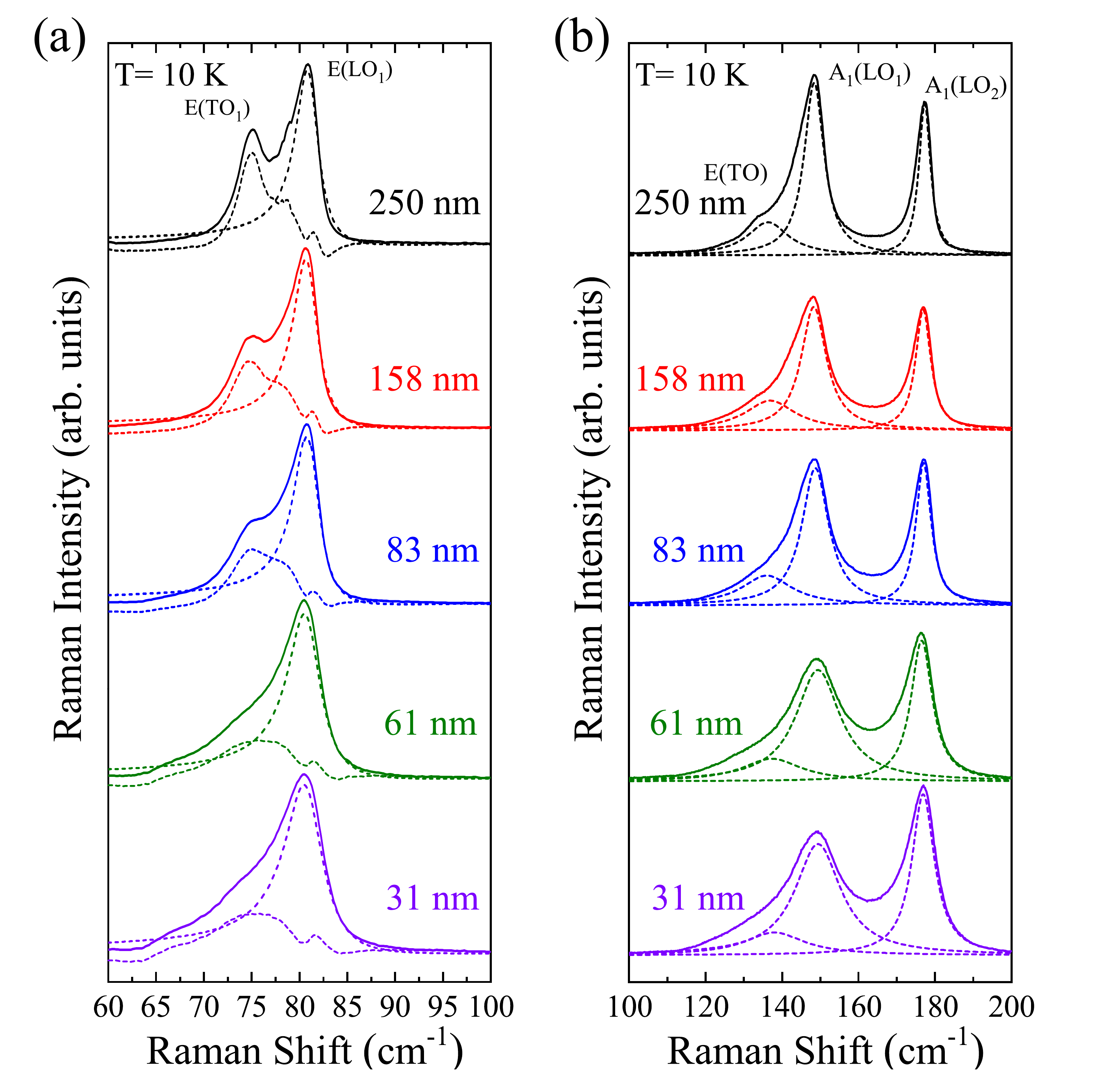}
		\caption{Raman spectra at 10 K for the five sizes of nanoparticles of the (a) E(TO$_1$) and E(LO$_1$) phonons and the (b)  A$_1$(LO$_1$) and A$_1$(LO$_2$) phonons.}
		\label{fig:2}
	\end{center}
\end{figure}

Figure \ref{fig:2}(a) shows the Raman spectra of the E(TO$_1$) and E(LO$_1$) phonons at 75 cm$^{-1}$ and 81 cm$^{-1}$  respectively for the five sizes. The temperature dependence of the E(TO$_1$) phonon is responsible for the variation of the dielectric constant and is a potential candidate to be the soft mode driving the ferroelectric transition \cite{Kamba2007,Lobo2007}. It has been observed in infrared spectroscopy that, compared to that of the E(LO$_1$) phonon, the intensity of the  E(TO$_1$) phonon decreases for sizes smaller than 83 nm \cite{Chen2010}. This behavior has been ascribed to the depolarization of the nanoparticles. This size-dependent intensity can be expressed as a function of the inverse nanoparticle diameter. The critical size below which the nanoparticles become paraelectric can be determined using this relation and the relative intensity of the E(TO$_1$) and E(LO$_1$) \cite{Chen2010}. We find a critical size of 12 nm in agreement with previous measurements around 10 nm \cite{Selbach2007}.

\begin{figure}[h]
	\begin{center}
		\includegraphics[width=8.7cm]{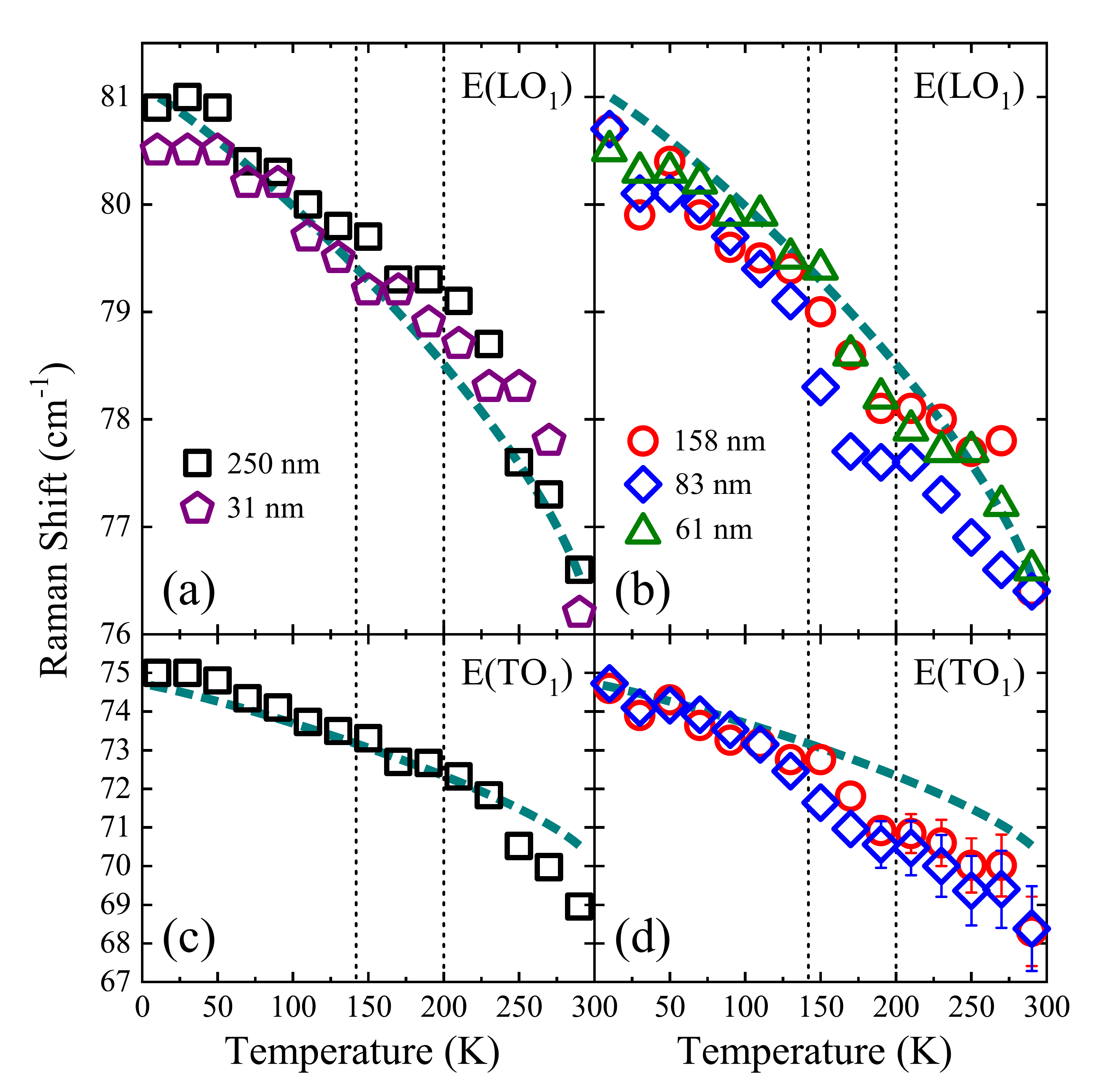}
		\caption{Temperature dependence of the frequency of the  E(LO$_1$) phonon mode (a) for a size of 250 nm and 31 nm and (b) for a size of 158 nm, 83 nm and 61 nm. Temperature dependence of the frequency of the  E(TO$_1$) phonon mode (c) for a size of 250 nm and (d) for a size of 158 nm and 83 nm. Dashed lines correspond to the behavior in the bulk. Vertical dashed lines represent the range where the surface expansion is expected. The error bars are visible for some points because they exceed the size of the symbols.}
		\label{fig:3}
	\end{center}
\end{figure}

The frequencies of the two phonons are reported in Fig. \ref{fig:3}. To have access to the frequency of the E(TO$_1$) mode, the E(LO$_1$) is subtracted by a Fano shape. Notice that the E(TO$_1$) mode is too low in intensity to be studied as a function of the temperature for the sizes below 83 nm. In Fig. \ref{fig:3}(a) and (c), the temperature dependence of the frequencies for the 250 nm is similar to the one observed in the bulk. For 158 nm, both mode frequencies present a jump at around 140 K followed by a plateau at around 220K (Fig. \ref{fig:3}(b) and (d)).  This behavior is even more pronounced for 83 nm. At 61 nm, the presence of the jump and the plateau are less obvious in the frequencies of E(LO$_1$). They disappear for a size of 31 nm for which the temperature dependence of the frequency is close to the one at 250 nm. The surface phase transition no longer plays a role for a size of 31 nm. 

\begin{figure}[h]
	\begin{center}
		\includegraphics[width=8.7cm]{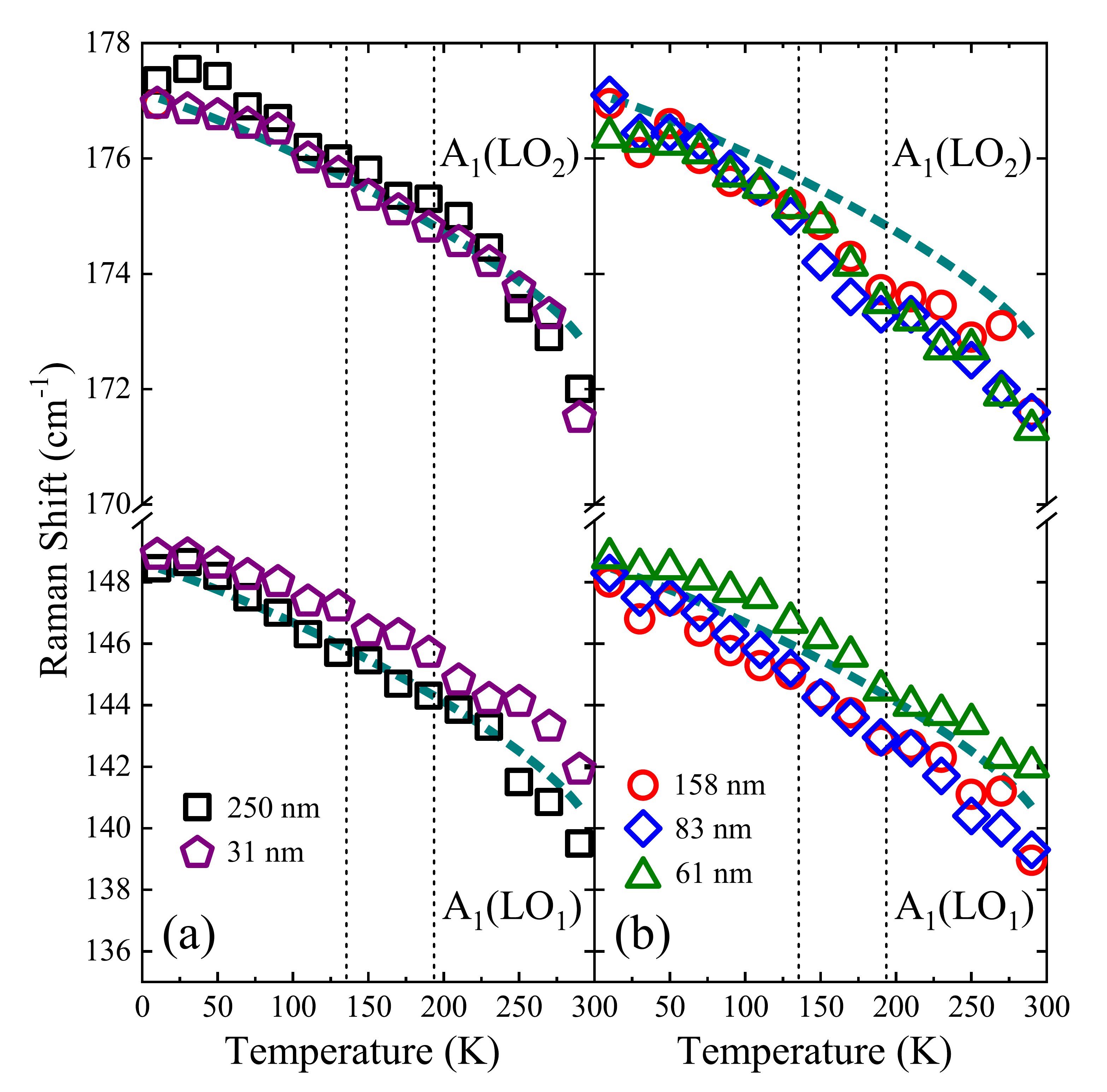}
		\caption{Temperature dependence of the frequency of the A$_1$(LO$_1$) and A$_1$(LO$_2$) phonons mode (a) for a size of 250 nm and 31 nm and (b) for a size of 158 nm, 83 nm and 61 nm. Dashed lines correspond to the behavior in the bulk. Vertical dashed lines represent the range where the surface expansion is expected.}
		\label{fig:4}
	\end{center}
\end{figure}

Figure \ref{fig:2}(b) shows the Raman spectra of the  A$_1$(LO$_1$) and A$_1$(LO$_2$) phonons at 148 cm$^{-1}$ and 177 cm$^{-1}$  respectively for the five sizes. The temperature dependencies of their frequency are plotted in Fig. \ref{fig:4}. The frequencies of the A$_1$(LO$_2$) mode present the same size dependent anomalies (jump and plateau) as the ones measured for the E(LO$_1$) mode. %Similarly to magnetic excitations, our results show that the surface has a significant impact on the lattice for a size approaching the cycloid periodicity. 
Notice that the other phonons, such as the A$_1$(LO$_2$) mode, exhibit a classical behavior for each size corresponding to the thermal expansion of the lattice.

These measurements show that there is an unusual trend associated with particular sizes. First of all, we note the disappearance of the fingerprint of the cycloidal spin excitations between the sizes of 83 nm and 61 nm. Moreover, the frequency of the magnetic excitations and the frequency of the E(LO$_1$), E(TO$_1$) and A$_1$(LO$_1$) phonons present identical anomalies at the same temperatures related to the surface expansion. In addition, the anomalies are enhanced for 83 nm and disappeared for 31 nm. We remark that the only comparable physical scale to these peculiar sizes is the periodicity of the cycloid (62 nm).

The first question that arises is the critical size of the nanoparticles below which the spiral of spins is either strongly modified or suppressed. From an experimental point of view, previous Mössbauer spectroscopy measurements have detected the cycloid state for a size of 54 nm \cite{Landers2014}. The modification of the cycloid can come from the uncompensated spins at the surface \cite{Park2007} and the magnetic surface anisotropies. However magnetic surface anisotropies can dominate magnetic properties only in very small nanoparticles \cite{DeSousa2019} and the cycloid should adapt itself. The disappearance of the Raman signature of the spin excitations below 83 nm in this work pleads in favor of the suppression of the cycloid between 83 and 61 nm. This possibility cannot be totally ruled out. One reason for the vanishing of the Raman peaks for 61 and 31 nm might be inhomogeneous broadening. Previous theoretical work from Ref. \onlinecite{DeSousa2019} shows that the cycloid wavevector is highly sensitive to size fluctuations for these small sizes. Inhomogeneous broadening is expected to be large, and it would greatly reduce the sensitivity to Raman. From this work, inhomogeneous broadening of magnon peaks might prevent the detection of Raman peaks associated to the cycloid. Other techniques such as Mössbauer and neutron scattering must be used to answer this question definitively.

Our measurements show the impact of the surface expansion on magnetic and lattice modes. The link between both types of excitations is explained by their coupling through magnetic interactions \cite{Matsuda2012, Fishman2013} and magneto-striction \cite{Arnold2009,Lee2015}. Notice that the surface expansion is only related to lattice properties. Indeed, it should have affected the 31 nm more than all the others sizes because the surface-to-volume ratio is the largest for the 31 nm particle. It has been shown that the ferromagnetic moment and the magneto-electric coupling are enhanced for a size of BiFeO$_3$ nanoparticles approaching the periodicity of the cycloid. Around this size, the rotation angle of the FeO$_6$ octahedra (Fig. \ref{fig:0}(a)) increases due to lattice strain. We recall that this rotation angle determines the local ferromagnetic moment through Dzialoshinskii-Moriya interaction. Since this local ferromagnetic moment is no longer cancelled out by the cycloid at the surface, it leads to an anomalous increase of the magnetization at the size of 61 nm \cite{Huang2013}. The behavior of the frequency of phonons and magnons could be a signature of the interplay between this additional ferromagnetic moment and the surface.

Such a coupling between surface, lattice and magnetic properties in nanoparticles makes BiFeO$_3$ a promising compound for applications wherein the functionalization of surfaces can improve performance, for example. It would then be possible to envisage a multiferroic nanoparticle with a very strong magneto-electric effect where the electric and magnetic properties, not only of the nanoparticle itself, but also of any material deposited onto its surface could be simultaneously tuned.

In conclusion, we have investigated both the magnetic and structural modes of BiFeO$_3$ as a function of particle size. We have found compelling evidence for strong frequency anomalies of phonons and magnons at a size of 83 nm associated with an expansion of the surface. This work reveals a cross- correlation between the lattice, the magnetism, and the surfaces in BiFeO$_3$ nanoparticles at a size close to the cycloid periodicity.\\\\

%%%%%%%%%%%%%%%%%%%%%%%%%%%%%%%%%%%%%%%%%%%%%%%%%%%%%%%%%%%%%%%%%%%%%
\par
	Work on this class of materials in SSW’s laboratory is supported by the U.S. National Science Foundation under Grant No. CHE-1807640.

%\subsection{References}

\end{document}